\documentstyle[12pt,psfig]{article}
\begin{document}
\title{
{\small COLLECTIVE EXCITATIONS OF A PERIODIC BOSE
CONDENSATE\\ IN THE 
WANNIER REPRESENTATION}}

\author{\small M. L. Chiofalo, M. Polini  and M. P. Tosi$^*$\\
\small\it Istituto Nazionale di Fisica
della Materia and Classe di Scienze,\\
\small\it  Scuola Normale Superiore, I-56126 Pisa, Italy\\
}

\date{}

\maketitle
\begin{abstract}

We study the dispersion relation of the excitations of a dilute
Bose-Einstein
condensate confined in a periodic optical potential and its Bloch
oscillations in an accelerated
frame. The problem is reduced to one-dimensionality through a
renormalization of the s-wave
scattering length and the solution of the Bogolubov - de Gennes
equations is formulated in terms of the appropriate Wannier 
functions.
Some exact properties of a periodic one-dimensional 
condensate are easily demonstrated:
(i) the lowest band at positive energy 
refers to phase modulations of the
condensate and has a linear dispersion relation near the Brillouin
zone centre; (ii) the higher bands arise from the superposition of
localized excitations with definite phase relationships; 
and (iii) the wavenumber-dependent current 
under a constant force in the semiclassical transport
regime vanishes at the zone boundaries.  
Early results by J. C. Slater 
[Phys. Rev. {\bf 87}, 807 (1952)] 
on a soluble problem in electron energy
bands are used to specify
the conditions under which the Wannier functions
may be approximated
by on-site tight-binding orbitals of 
harmonic-oscillator form. In this approximation the connections
between the low-lying excitations in a lattice and those in a
harmonic well are easily visualized. 
Analytic results are obtained in the tight-binding 
scheme and
are illustrated with
simple numerical calculations for the dispersion relation and
semiclassical transport in the lowest energy band, 
at values of  the system
parameters which are
relevant to experiment.
\end{abstract}


\vfill

\noindent ${}^*$ Corresponding author. 
Tel.: +39-050-509275; fax: +39 50 563513; e-mail:
{\tt tosim@bib.sns.it}.

\newpage

\section{Introduction}

The dynamics of cold atomic vapours trapped in a periodic
external potential, as
created by means of a far-detuned standing 
wave of light, has become an
active area of
research. Studies of ultracold atoms in such a confinement have revealed a
number of
behaviours which are well known in solid state physics: Wannier-Stark
ladders [1, 2], Bloch
oscillations [3, 4] and Landau-Zener tunneling [5]. More recently, the
observation of
interference between macroscopic quantum states of Bose-Einstein-condensed
vapours of ${}^{87}$Rb
atoms confined in a vertical array of optical traps has
been reported [6].
Interference between the condensates in different lattice wells manifests
itself in falling drops,
which are emitted with a period determined by the Bloch oscillations 
induced by the gravitational field and are interpreted as
coherent matter-wave pulses.
	
The properties of a Bose-Einstein condensate (BEC) in a regular
array of optical wells
have been studied theoretically and numerically by a number of authors [7 -
13]. Especially
relevant for our present purposes is the work on an interacting BEC in an
infinitely extended
lattice potential by Berg-S\o rensen and M\o lmer 
[7], who demonstrated band
structure and Bloch
oscillations by solving numerically the Bogolubov - de Gennes equations
for the excitations
and the time-dependent Gross-Pitaevskii equation for motion under a
constant applied force.
They also gave analytic results for the excitation spectrum in the limit
where the periodic
potential can be treated as a perturbation.

From solid state theory a tight-binding (TB) approach offers the
possibility of an analytic
study of the same phenomena in the opposite limit, when the lattice
potential is so strong
compared with all other energy scales as to allow the use of on-site
orbitals for the approximate
construction of the Bloch states. In fact, the proper theoretical tools are
the so-called Wannier
functions, which are defined through the Fourier transform of the Bloch
eigenfunctions in the
momentum representation. In the particular case of one-dimensional (1D)
motion, the energy
bands are non-degenerate and the states in any given one of them are
described from a single
Wannier function, by repeating it at each lattice site with the appropriate
phase relationship
between different sites. The TB scheme may then be realized by replacing
the Wannier function
by an on-site "atomic" orbital, under suitable limiting conditions which
may be valid for the low-lying energy bands.

While the Wannier functions for motion in a generic periodic
potential are usually not easy
to calculate, it turns out that for the specific problem of 1D periodicity
posed by the
confinement of a BEC in an array of optical wells one can take
advantage of the results contained in a classic paper
by Slater [14] on a soluble problem in energy bands for electrons. Slater's
work also discusses
the conditions under which the Wannier functions for the low-lying energy
bands can be
accurately approximated by on-site orbitals, which in this case are the
harmonic-oscillator wave
functions associated with the parabola describing the bottom of each well.

The plan of the paper is as follows. In section 2 we first
model a dilute BEC
confined in a periodic optical potential as a 1D dynamical system via
inclusion of the
transverse size into a renormalized contact interaction. We then formulate
the problem of 1D
band structure in terms of Wannier functions, the only conceptual
difference from the
analogous problem of 1D motion of electrons in a periodic potential being
that for condensed bosons there
is no gap between 
the condensate ground state and the excitation spectrum. In the same
section we also
formulate the problem of Bloch oscillations of the condensate 
in a single band in terms of the Fourier transform
of the corresponding Wannier function.

The reduction of the general formalism to the TB approximation is
effected in section 3, after
recalling Slater's conditions for its validity and extending them to an
interacting BEC. The
TB scheme displays in an explicit
manner how the energy
levels of a BEC in a single well are shifted and spread into bands by the
periodic replica of
the well potential. We give special attention to (i) 
the lowest-lying energy
band of the BEC, which
contains its wave-like phase excitations, and (ii)
the connection that the TB approximation exhibits between the
excitations in the next two higher bands and the dipole 
(``sloshing'') and quadrupole excitations in an effectively 1D
harmonic well. The discussion suggests how excitations in these two
bands may be observed. We conclude this section by simple numerical
illustrations of the dispersion relation in the lowest band
and of the current induced
by a constant external
force, for values of the system parameters which are relevant to the
experiments of Anderson
and Kasevich [6]. Finally, section 
4 ends the paper with some concluding remarks.

\section{Band structure and Bloch oscillations of a BEC
from Wannier orbitals}

We consider a dilute vapour of atoms of mass $m$  
in a Bose-condensed state at zero
temperature, which interact via elastic collisions described by an s-wave
scattering length $a$  and
are confined in the optical lattice created by a laser field. The lattice
can be modelled as a
transverse Gaussian potential times a periodic potential along the $z$
direction, namely the confining potential is
\begin{equation}
\label{perpo}
U_l(r,z)=U_0 \exp(-r^2/r_{lb}^2) \sin^2(kz)\; .
\end{equation}
Here $U_0$  
is the well depth, scaling linearly with the intensity of the laser
beam, $r_{lb}$  is the
transverse size of the beam and $k=2\pi/\lambda$
its wavenumber. The
lattice period is $d=\lambda/2$.

In this work we reduce the above problem in cylindrical symmetry to
an effective 1D
problem. To this end we freeze the transverse motions of the BEC and
renormalize the
interactions by a method proposed by Jackson {\it et al.} [15] to treat 
an elongated BEC, in which the coherence
length as introduced by Baym and Pethick [16] 
is much larger than the transverse size. This reduction will be
inapplicable, however, to high-energy excitations, when the excitation
energy becomes comparable with the transverse confinement energy and
transverse motions can therefore be induced.
 
Under the conditions indicated just above, the transverse
confinement modifies the mean-field interactions by a factor depending on
the axial density $\sigma(z)$, 
the factor being  $\int dxdy\ |G(x,y;\sigma(z))|^4$  where $G$   
is the transverse part of the wave function
in the $xy$  plane. 
Using a harmonic approximation for the radial part of the
optical potential, we
obtain in the present problem an effective scattering length per unit area
given by
\begin{equation}	
\label{asc}
\tilde{a}={an_0\over r_{lb} d}\sqrt{{2U_0\over E_R}}\; ,
\end{equation}
where $n_0$  is the number of atoms per lattice well and 
$E_R=h^2/(md^2)$. Notice
that the energy $E_R$  differs by a factor eight from the recoil 
energy $E_{recoil}=\hbar^2 k^2/2m$. A similar
renormalization
of $U_0$  in Eq. (\ref{perpo}) is negligible, since $r_{lb}\gg\lambda$.

The effective 1D Hamiltonian $H$  that we treat thus is
\begin{equation}
\label{hamiltonian}
H=\int dz \psi^\dagger (z)\Lambda(z)\psi^{}(z)+{1\over 2}g\int dz
\psi^\dagger (z)\psi^\dagger (z)\psi^{} (z)\psi^{} (z)
\end{equation}
where $\psi(z)$  is the field operator, $g=4\pi\hbar^2\tilde{a}/m$
and  $\Lambda(z)=-(\hbar^2/2m)\nabla^2+U_p(z)-\mu$, with
$\mu$  the chemical potential and $U_p(z)=U_0\sin^2(kz)$.
In the following all energies are referred to the bottom 
of the lattice potential.

\subsection{Lattice symmetry and Bose statistics}
Let us summarily go through the standard Bogolubov procedure [17] for
treating the boson
Hamiltonian (\ref{hamiltonian}). 
We make the Ansatz $\psi(z)=\phi(z)+\tilde{\psi}(z)$, 
which separates the coherent 
condensate function $\phi(z)$ 
from the fluctuating part $\tilde{\psi}(z)$  in the field operator. Upon
substituting this Ansatz in Eq. (\ref{hamiltonian}) 
and requiring that linear terms 
in $\tilde{\psi}(z)$ vanish, one
obtains the Gross-Pitaevskii equation for $\phi(z)$:
\begin{equation}
\label{gpe}
\left[\Lambda(z)+g\phi^2(z)\right]\phi(z)=0\; .
\end{equation}
We are taking $\phi(z)$ as real without loss of generality, 
with normalization to unity over the lattice period. 

To quadratic terms the Hamiltonian
in the fluctuation operators is diagonalized by the canonical
transformation
\begin{eqnarray}
\tilde{\psi}(z)&=&\sum_\alpha\left[u^{}_\alpha(z)a^{}_\alpha-
v_\alpha(z)a_\alpha^\dagger\right]\; ,\nonumber\\
\tilde{\psi}^\dagger(z)&=&\sum_\alpha\left[u^*_\alpha(z)a^\dagger_\alpha-
v^{*}_\alpha(z)a_\alpha\right]\; .\label{bogo}
\end{eqnarray}
In Eq. (\ref{bogo}) the suffix $\alpha$  
denotes an appropriate set of quantum numbers and
the functions $\{u_\alpha,v_\alpha\}$
satisfy the Bogolubov - de Gennes equations,
\begin{eqnarray}
L(z)u^{}_\alpha(z)-H_I(z)v^{}_\alpha(z)&=&
E_\alpha u^{}_\alpha(z)\; ,\nonumber\\
L(z)v^{}_\alpha(z)-H_I(z)u^{}_\alpha(z)&=&
-E_\alpha v^{}_\alpha(z)
\; .\label{bogoeq}
\end{eqnarray}
Here, $H_I(z)=g\phi^2(z)$, 
$L(z)=\Lambda(z)+2H_I(z)$ and the $E_\alpha$'s are
the (real) excitation energies that we aim to determine. We explicitly
point out that with this form of $H_I(z)$ the interactions enter the
determination of the excited-state wave functions only through the
density distribution of the condensate ground state. The same
assumption has been made in dealing with the excitations of
condensates in harmonic traps [18] and has led to good agreement with
experiment for the eigenfrequencies of the quadrupolar surface modes.

For a BEC in a 1D periodic potential the
quantum numbers $\alpha$  are the
band index $n$ and the lattice quasi-momentum $q$. 
From symmetry under time
reversal we have
that, if $\{u_{nq},v_{nq}\}$ is a solution of Eqs. (\ref{bogoeq}) 
corresponding to energy $E_{nq}$ , then $\{v_{nq},u_{nq}\}$
is a solution corresponding to energy $-E_{nq}$ [19,20]. 
We may for convenience call these  the particle-like
and the hole-like solution, respectively, and limit ourselves to calculate
spectrum at positive energy. As for the homogeneous Bose fluid we expect
gaplessness,
since in its ground state the whole condensate
is in a single
energy level placed at the chemical potential. Therefore, the excitations
of lowest energy
(referred to the chemical potential) are phonons, i.e. they have a linear
dispersion relation. 

It is convenient 
to introduce solutions $Z_{nq\pm}(z)$
which have definite symmetry
under time reversal [7], by setting $Z_{nq\pm}(z)=
\left[u_{nq}(z)\pm v_{nq}(z)\right]/\sqrt{2}$. 
From equations (\ref{bogoeq}) we have
\begin{equation}
L_-(z)Z_{nq+}(z)=E_{nq}Z_{nq-}(z) \label{bogoeqZa}
\end{equation}
and
\begin{equation}
L_+(z)Z_{nq-}(z)=E_{nq}Z_{nq+}(z)\; , \label{bogoeqZb}	
\end{equation}
where $L_\pm (z)=L(z)\pm H_I(z)$. 
Hence,
\begin{equation}
\label{bogoeqE2}
L_\pm (z)L_\mp (z)Z_{nq\pm}(z)=E_{nq}^2Z_{nq\pm}(z)\; .
\end{equation}
This equation shows that the functions $Z_{nq+}(z)$  and
$Z_{nq-}(z)$ are
the right and left eigenvectors of the non-hermitean product
operator $L_+ (z)L_- (z)$.

The ground state of the condensate can be taken at $q=0$ in the 
lowest band ($n=0$, say) and 
special care is needed in handling the functions $Z_{00\pm}(z)$, 
in order to
ensure gaplessness. This point is discussed for a periodic condensate 
in Appendix A, following the arguments given in Refs. [19] and [20].
Here we only need to note that an evaluation of the functions 
$Z_{nq-}(z)$ is not necessary for the determination of the 
excitation energy
eigenvalues. These are calculated by solving 
Eq. (\ref{bogoeqE2}) for $Z_{nq+}(z)$, which reads
\begin{equation}
\label{bogoeqE2+}
L_+ (z)L_- (z)Z_{nq+}(z)=E_{nq}^2Z_{nq+}(z)\; .
\end{equation}
This equation applies also to the state $Z_{00+}(z)$ at zero energy
(relative to the chemical potential).
Of course, the function $Z_{00+}(z)$ is simply proportional to the
function $\phi(z)$ introduced in Eq. (\ref{gpe}). The solutions of
Eq. (\ref{bogoeqE2+}) must have the Bloch symmetry, which imposes
definite $q$-dependent phase relationships between different wells.

\subsection{Expansion in terms of Wannier functions}
It is well known from
solid state theory [21] that any Bloch orbital can be expressed as
a superposition of the so-called Wannier
functions centred on the lattice sites. We thus write
\begin{equation}
\label{wannier}
Z_{nq\pm}(z)={1\over \sqrt{N}}\sum_l\exp (ilqd)w_{n\pm}(z-ld)
\end{equation}
where the index $l$ runs over all $N$ sites of the lattice. 
Of course, the Wannier functions $w_{n\pm}(z)$ have the same parity
under time reversal as the corresponding Bloch orbitals 
$Z_{nq\pm}(z)$.
We may also recall that the Wannier
functions form a complete orthonormal set, i.e. they satisfy the
relation $\int dz\; w^*_{n\pm}(z-ld) w_{n'\pm}(z-l'd)=
\delta_{nn'}\delta_{ll'}$. Orthogonality between different wells
implies oscillatory tails, falling off more rapidly as overlap decreases.

Eq. (\ref{wannier}) expresses a Fourier-transform relationship between
Bloch orbitals and Wannier functions, embodying the Bloch
translational symmetry. The $q$-dependent Bloch orbitals emphasize the
extended nature of the lattice states and the periodicity of their
energy eigenvalues in momentum space. The representation in terms of
Wannier functions emphasizes instead that the lattice is a periodic
assembly of identical wells. The Bloch symmetry of the Fourier series
in Eq. (\ref{wannier}) is ensured by the fact that the Fourier
coefficients do not involve a separate dependence on position $z$ and
on lattice site index $l$.  

In the case of a BEC in a lattice,
a general consequence
of the Wannier-function description is immediately evident from 
Eq. (\ref{wannier}).
Since the ground state is at the chemical potential and has been taken
to lie 
at $q=0$ in the bottom of the lowest band, 
the excitations in this band are wave-like modulations 
of the condensate phase and
their energy vanishes linearly in the limit $q\to 0$.
This fact can be traced back to the existence of 
a Goldstone phonon - a collective mode without restoring force
[19] resulting from spontaneous symmetry breaking (see 
the discussion in Appendix A). 

Similarly, for the higher excitation bands Eq. (\ref{wannier}) shows
that each of them is a superposition of localized excitations
(described by $w_{n+}(z)$), which is constructed by imposing definite
phase relationships between the various sites as dictated by the
translational lattice symmetry. This theorem is valid, of course,
within the Bogolubov approach and has been derived through 
the reduction to one-dimensionality. 

In practice, the calculation of Wannier
functions is in general not an easy task. In the limit in which the
Wannier functions become almost confined inside each lattice well
(which may be true for low--lying states), 
they can be expanded in a limited number of ``atomic''
orbitals. We
shall discuss in the next section the conditions under which one may
replace the Wannier
functions for a BEC in a periodic optical potential by single-well 
orbitals and thereby obtain explicit solutions in
the TB limit.

\subsection{Semiclassical transport in band states}
We proceed to express through the Wannier functions the current carried by
a periodic Bose-condensed 
system subject to a constant force $F$ 
in the semiclassical regime.
We restrict
ourselves to Bloch oscillations in the lowest band, 
described by the
Wannier function $w_{0+}(z)$.
Through Eq. (\ref{wannier}) taken at $q=0$, 
this function is immediately related to the ground state 
$\phi(z)$  of the condensate.

The quantity of interest is the expectation value $\bar{p}(t)$ of the
momentum operator $p=-i\hbar d/dz$
in the state $Z_{0q(t)+}(z)$, where $q(t)=Ft/\hbar$. 
We introduce the Fourier transform
$f_0(k)$ of $w_{0+}(z)$,
\begin{equation}
\label{FT}
f_0(k)={1\over \sqrt{2\pi}}\int_{-\infty}^{\infty} dz\; w_{0+}(z)
\exp (-ikz)\; .
\end{equation}
Hence,
\begin{equation}
\label{Z0q}
Z_{0q+}(z)={1\over \sqrt{2\pi N}}\sum_l\int_{-\infty}^{\infty} dk 
\; f_{0}(k)\exp [i(q-k)ld+ikz]
\end{equation}
and a straightforward calculation yields
\begin{equation}
\label{pmedio}
\bar{p}(q(t))={\hbar\sum_l\int_{-\infty}^{\infty} dk 
\; k|f_{0}(k)|^2\cos [ld(q(t)-k)]\over 
\sum_l \int_{-\infty}^{\infty} dk\; 
|f_{0}(k)|^2\cos [ld(q(t)-k)]}
\end{equation}
This is an exact expression, which holds independently 
of the confining
periodic potential and
of the interaction between the particles. 

It is easily seen from Eq. (\ref{pmedio}) that (i)
$\bar{p}(t)$ is a periodic function
of time with the Bloch
period $T_B=h/(Fd)$ and (ii) $\bar{p}(t)$ 
vanishes at the Brillouin zone edges, i.e. at   
$q=\pm\pi/d$ or $t=\pm T_B/2$.

\section{Tight-binding approximation}
The aim of the TB approximation is to relate 
the properties of the periodic system in a 
lattice potential to the solution of
the single (isolated) well problem. 
The results of the method are quantitative as long as the single-well
solutions at adjacent sites do not overlap much with each other. 
Namely, the validity of 
the TB approximation is limited by the condition that 
the band-width be smaller than the
height of the lattice potential.

Slater [14] has studied the Wannier functions 
of the "Mathieu problem" defined 
by the Schr\"odinger equation for non-interacting electrons moving 
in the periodic potential
$U_p(z)=U_0\sin^2(kz)$
of present interest. He showed that for the first five
low-lying energy bands the Wannier functions
are well approximated by Hermite
polynomials under the
condition $U_0>2E_R$, relaxing to $U_0>E_R/4$   
if only the lowest energy band is of interest.
Slater's argument is summarized in Appendix B and extended there to
an interacting
BEC. In essence the mean-field interactions may be viewed in the present
context as effecting
a renormalization in the height of the external potential, thereby
shifting Slater's thresholds by the amount 
\begin{equation}
\label{slaterineq}
{\Delta U_0}={\tilde{a}d E_R \over \pi}\; .
\end{equation}
If the threshold conditions are satisfied, 
the Wannier functions $w_{n+}(z)$ at positive
energy could be approximately replaced by the
eigenfunctions of the
harmonic approximation to the single-well potential 
$\overline{U}^{(s)}_p(z)=\overline{U}_0\sin^2(kz)$ 
with $z\in\left[-d/2,d/2\right]$ and 
$\overline{U}_0=U_0+\tilde{a} d E_R/\pi$.

\subsection{Band structure for a periodic BEC}
More generally, the evaluation of the energy bands for a periodic
condensate in the TB
approximation would require a preliminary solution of the on-site problem. 
To this end, 
we separate the on-site parts $L^{(s)}_\pm(z)$ out of the operators 
$L_\pm(z)$ entering Eq. (\ref{bogoeqE2}) by writing       
$L_\pm(z)=L^{(s)}_\pm(z)+\Delta L_\pm(z)$.
The detailed definitions of these operators are given in Appendix C. 
The on-site operators $L^{(s)}_\pm(z)$ 
determine the single-well orbitals $\varphi_{n\pm}(z)$ 
corresponding to single-well energy eigenvalues $\epsilon_n$, 
according to
\begin{eqnarray}
L^{(s)}_-(z)\varphi_{n+}(z)&=&\epsilon_{n}\varphi_{n-}(z)\; ,
\nonumber\\
L^{(s)}_+(z)\varphi_{n-}(z)&=&\epsilon_{n}\varphi_{n+}(z)\; .
\label{bogoeqs}
\end{eqnarray}		                   
Again, the condensate in a single well lies at the chemical potential
$\mu_s$ (see the definitions of $L^{(s)}_\pm(z)$ in Appendix C).
Conjugation between fluctuations in the particle 
number and in the phase must also
be taken care of, as shown in Refs. [19] and [20].

We define the overlap integrals
${\mathcal I}_n(l)=\int dz\; \varphi^*_{n-}(z)
\varphi^{}_{n+}(z-ld)$ and the interaction 
integrals ${\mathcal J}_n(l)=\int dz\; \varphi^*_{n-}(z)
\Delta\left[L_+(z)L_-(z)\right]\varphi^{}_{n+}(z-ld)$ (see the
definitions of the $\Delta L_\pm(z)$ operators given in Appendix C).
By exploiting Eqs. (\ref{bogoeqs})
one easily finds from Eq. (\ref{bogoeqE2}) the dispersion relation
\begin{equation}
\label{bandgen}
E_{nq}^2=\epsilon_n^2+{\sum_l {\mathcal J}_n(l)\exp (ildq)\over 
\sum_l {\mathcal I}_n(l)\exp (ildq)}\; .
\end{equation}
An adiabatic argument regarding the switching-on of the 
(spatial-symmetry conserving) mean--field interactions shows
that the product
$\varphi^*_{n-}(z)\varphi_{n+}(z')$ is even under space inversion. 
Both series in Eq. (\ref{bandgen}) are therefore real quantities.

Eq. (\ref{bandgen}) displays the two main changes induced by lattice
periodicity on the single-well energy levels, namely (i) a shift of
each level relative to the single-well value $\epsilon_n$ and
(ii) the broadening of the level into a band. In particular, for the
lowest band the shift corresponds to a renormalization of the chemical
potential from the single-well value $\mu_s$ to the value $\mu$, which
has to be evaluated self-consistently from the condition
$E_{00}=\mu$. The resulting gaplessness introduces in Eq. (\ref{bandgen})
for $n=0$ 
the linear, acoustic-phonon-like dispersion of the excitation energies
in the long wavelength limit $q\to 0$.

We can now see how the
excitations in the bands with $n=1$ and $n=2$ may be
related to single well excitations. We make use of
the present replacement of Wannier functions by on-site orbitals, which
are approximately described by the eigenfunctions of the 1D harmonic
oscillator. 
Evidently, this allows us to identify 
the on-site excitation leading to the
$n=1$ band with the dipolar sloshing mode of the condensed particles
within the well. The construction of Bloch orbitals from these
localized sloshing motions simply amounts to fixing, for each value of
the wavenumber $q$, their relative phases according to the lattice
translational symmetry (see the discussion already given near the end
of section 2.2). Similarly, the excitations in the $n=2$ band may be
viewed as resulting from localized, single-well
quadrupolar modes with relative phases depending on $q$ as imposed by
lattice symmetry. Of course, the approximate validity of such a
visualization of the collective low-lying modes of a periodic
BEC is subject not only to conditions relating to effective
one-dimensionality in a mean-field approach, but also to the restrictions
associated with the TB scheme.

The above discussion also suggests how these collective band modes
may become accesible to experimental or computational observation. 
Clearly, at $q=0$ the dipolar mode can be excited by a rigid shaking
of the lattice and the quadrupolar one by a modulation of the width 
of all the wells. 
Excitation of dipolar and quadrupolar band modes at finite $q$
would require similar probes at given wavelengths.

\subsection{Dispersion relation and 
Bloch oscillations in the lowest band}

We conclude this section by presenting some illustrative numerical
results for the lowest band in the TB
approximation. According to Eq. (\ref{slaterineq}), 
in the case of repulsive interactions the inequality 
$U_0>(0.25+\tilde{a}d/\pi)E_R$
must be satisfied for the appropriate Wannier function 
to be approximated
by the Gaussian function
\begin{equation}
\label{gauss}
\varphi_{0+}(z)=
{1\over(\sqrt{\pi}\sigma)^{1/2}}\exp (-z^2/2\sigma^2)\; .
\end{equation}
The width $\sigma$  
of this Gaussian is determined in a first approximation by the
curvature of the
bottom of the single well. A more accurate estimate of $\sigma$, 
which takes account of
the interactions
between the particles in the BEC, is given in Appendix C.

In fact, for the calculation of the dispersion relation it is 
more convenient to 
revert back to Eq. (\ref{bogoeqE2+}) 
and directly evaluate
the integrals
\begin{equation}
K(l)=\int dz\;\varphi_{0+}(z)
\left[L_+ (z)L_- (z)\right]\varphi^{}_{0+}(z-ld)
\end{equation} 
and $I(l)=\int dz\;\varphi_{0+}(z)
\varphi^{}_{0+}(z-ld)$. This procedure has the advantage of involving 
only the known symmetric function $\varphi_{0+}(z)$ given in 
Eq. (\ref{gauss}). 
Since $I_0(0)\simeq 0.99$,
we obtain
\begin{equation}
\label{bandlow}
E_{0q}^2=\mu^2+2 K(1)\left[\cos(qd)-1\right]\; 
\end{equation}
where $\mu^2=K(0)+2K(1)$. 
We recall that the energies in Eq. 
(\ref{bandlow}) are referred to the bottom of the lattice 
potential.

We have
calculated the integrals
entering Eq. (\ref{bandlow}) 
with the wave function in Eq. (\ref{gauss}) for 
$U_0/E_R=0.25$, which corresponds to a value of $\sigma/d= 0.27$, 
in the case
$a=110\; a_0$, $r_{lb}=80\;\mu m$ and $n_0=10^3$. These values are 
relevant to the experiment of Anderson and Kasevich [6]. 
The results are
$K(1)\simeq -4\cdot 10^{-4} \; E_R^2$
and $\mu\simeq 0.1\; E_R$,
yielding the value $c_s\simeq 0.02 \; E_{R} d/\hbar$
for the speed of phonon-like excitations. Figure 1
shows
the corresponding dispersion curve over the whole Brillouin zone.

\begin{figure}
\centerline{\psfig{file=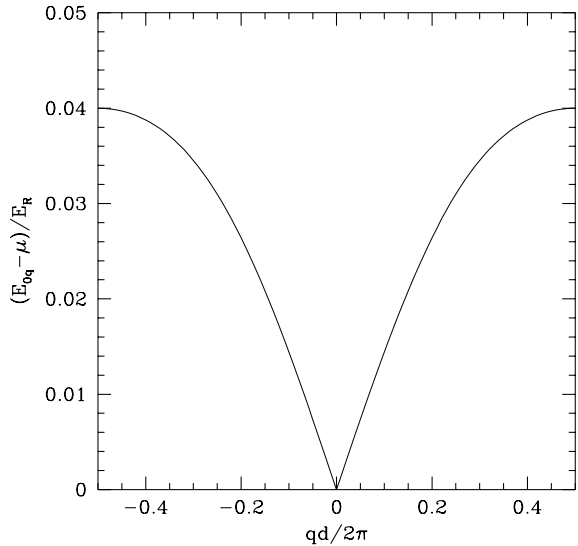,height=8cm,width=8cm}}
\caption{
An example of the dispersion relation of phase excitations in a
BEC confined in an
optical potential. We plot the lowest-lying energy band 
as a function of the wave number $q$ (in units of
$2\pi/d$) within the 1D
Brillouin zone, for values of the system parameters given by  
$a=110\; a_0$, $r_{lb}=80\;\mu m$, $n_0=10^3$, 
$U_0/E_R=0.25$ and $\sigma/d=0.27$.
}
\end{figure}

Using Eq. (\ref{gauss}) in Eqs. (\ref{FT}) and (\ref{pmedio}) 
we find the following expression
for the current
induced by a constant external force in states of the lowest band,
\begin{equation}
\label{pmedioTB}
\bar{p}(q(t))={\hbar d\over \sigma^2}{\sum_{l=1}^{\infty} l\; 
\sin (lqd)\exp\left[-\left (ld/2\sigma\right)^2\right]
\over 
1+2\sum_{l=1}^{\infty}  \cos
(lqd)\exp \left[-\left (ld/2\sigma\right)^2\right]}\; .
\end{equation}
In practice, in the TB limit the 
leading term in the numerator is dominant
and the denominator
can be replaced by unity. Namely,
\begin{equation}
\label{pmediofin}
\bar{p}(q)\simeq{\hbar d\over \sigma^2}\sin (qd)\exp 
\left[-\left(d^2/4\sigma^2\right)\right]\; .
\end{equation}
Figure 2 shows the behaviour of $\bar{p}(q)$    
over the Brillouin zone for
various values of the
parameter $\alpha=U_0/2E_R$ in the
case 
$a=110 a_0$, $r_{lb}=80\;\mu m$ and $n_0=10^3$.

\begin{figure}
\centerline{\psfig{file=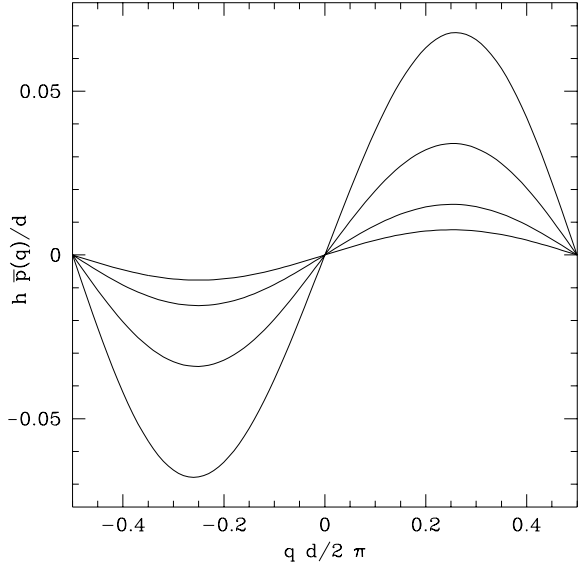,height=8cm,width=8cm}}
\caption{
Average momentum of the BEC 
in the lowest-lying energy band, in units of $h/d$, as a function
of the wave
number inside the 1D Brillouin zone, for the case
$a=110\; a_0$, $r_{lb}=80\;\mu m$, $n_0=10^3$ and
for various values of the parameter
$U_0/E_R$. From top to bottom on the RHS of the drawing:
$U_0/E_R=0.25,\; 0.4,\; 0.8$ and $1.2$.
}
\end{figure}

\section{Concluding remarks}
In summary, and as already discussed in the work of Berg-S\o rensen 
and
M\o lmer [7], the
properties which derive from lattice symmetry
for the wave functions and the energy
eigenvalues of
electrons moving in a periodic 1D potential also hold for a 
Bose-condensed
system. However,
Bose statistics and spontaneous 
symmetry breaking imply that there is no gap between the 
condensate ground state and the excitation 
spectrum.
This latter property can be met in 
the energy spectra of
electrons in semiconductors
only as a consequence of accidental degeneracies
induced by specific space symmetries, 
a well known instance being 
provided by graphite [22]. For a periodic Bose
condensate 
the gaplessness property, together with the quadratic structure of 
the energies as expressed
in Eqs. (\ref{bogoeqE2}) (see section 2.1),  
yields on general grounds the linear dispersion
relation $E_{0q}=c_s (|qd|)+O(q^2d^2)$ for the lowest energy band.

We have seen that the specific form of the optical potential acting
on a BEC from a laser
field admits two important simplifications in the 
band-structure problem.
Firstly, the problem for the low-lying bands
can be mapped into a one-dimensional one by accounting for the 
transverse
cross-section of the
BEC through a renormalization of the s-wave scattering length. 
This mapping
has been shown
to be useful in relation to the experiments performed by 
Anderson and Kasevich [12].
Secondly, the 1D problem away from the weak-confinement limit 
is most naturally formulated in terms of the Wannier 
functions, which have
already been studied for the Mathieu problem by Slater [14]. 
The use of Wannier functions, as demonstrated in section 2.2, 
provides an exact 
formulation of the lattice problem and immediately shows
that (i) the lowest excitations of a condensate in a periodic potential
are wave-like phase modulations, and (ii) the higher bands arise from
the superposition of localized excitations with definite phase
relationships. 

An analytic
study of semiclassical transport in a periodic condensate
subject to a constant external force 
becomes straightforward with the use of Wannier functions.
The expectation value of the momentum operator is periodic with 
the Bloch period and vanishes at the Brillouin zone edges, as
shown in section 2.3.

The work of Slater [14]
allows one to specify the conditions
under which the Wannier functions for the low-lying   
bands of the BEC in an optical periodic potential, 
and in particular that for the 
lowest band containing its phase 
excitations, can
be accurately approximated by harmonic-oscillator wave functions. 
These functions are simply
described by the curvature of the on-site well as renormalized by the
interactions in the BEC and allow an explicit, though approximate,
connection to be made between band modes and single-well modes. 
They also allow very simple calculations on a periodic BEC, as illustrated
in section 3 for the dispersion relation in the lowest energy band 
and for the 
average wavenumber-dependent momentum in correspondence to a
particular set of system parameters which is relevant to experiment.

\bigskip

\noindent{\bf Acknowledgments}.
This work is supported by the Istituto Nazionale di Fisica della Materia
through the Advanced
Research Project on BEC. We thank Professor G. Iadonisi for bringing
Slater's work to our attention.

\appendix
\section{The zero-energy solutions for the lattice problem}
The equation obeyed by the wave function $Z_{0q+}(z)$ at $q=0$, 
that is $Z_{00+}(z)\propto\phi(z)$,  is
\begin{equation}
L_-(z)Z_{00+}(z)=0\; .
\label{Z00+}
\end{equation}
We have chosen $E_{00}=0$ relative to the chemical potential $\mu$ of the
condensate (see the definition of $\Lambda(z)$ immediately below Eq. 
(\ref{gpe})). 

In view of the macroscopic occupation of this quantum
state, one must allow for fluctuations in the number of particles and
for the accompanying shifts of the phase of the condensate. This is
accounted for by adding in the Hamiltonian a ``kinetic energy'' term 
$\alpha P^2/2$, where $P=\int dz Z_{00+}(z)[\tilde{\psi}^{}(z)+
\tilde{\psi}^\dagger (z)]$ is the number operator generating such a
``translation'' and $1/\alpha$ plays the role of a mass. The
canonically conjugate operator 
$Q=i\int dz Z_{00-}(z)[\tilde{\psi}^{}(z)-
\tilde{\psi}^\dagger (z)]$ is easily seen to be the phase operator.
It satisfies the appropriate commutation relation provided that 
the weighting function $Z_{00-}(z)$ obeys the equation
\begin{equation}
L_+(z)Z_{00-}(z)=\alpha Z_{00+}(z)\; ,
\label{Z00-}
\end{equation}
with the condition $\int dz Z_{00-}(z)Z_{00+}(z)=1/2$.
 
From a more formal viewpoint [19, 20], 
the Bogolubov equations (\ref{bogoeq})
at zero energy admit the eigenvector $\{Z_{00+},Z_{00+}\}$. A second
solution is provided by the eigenvector of the squared Bogolubov
operator with zero eigenvalue. This implies that the result of
applying the Bogolubov operator to the second
solution is proportional to $\{Z_{00+},Z_{00+}\}$ through a constant
to be determined, leading to Eq. (\ref{Z00-}).
In fact, upon differentiation of the Gross-Pitaevskii equation 
with respect to the total number of particles $n$ one is led again to Eq. 
(\ref{Z00-}), with $Z_{00-}(z)=2\sqrt{n}\partial(\sqrt{n}Z_{00+}(z))/
\partial n$ and $\alpha=2 n\partial\mu/\partial n$. Clearly, the
constant $\alpha$ is related to the change induced in the chemical
potential $\mu$ by a fluctuation in the number of particles.

\section{Wannier functions for low-lying bands}

For the 1D Mathieu problem (non-interacting particles moving in the
potential $U_p(z)=U_0\sin^2(kz))$, Slater [14] 
shows that the eigenfunctions $f_n(k)$  in the momentum
representation
obey the adimensional finite-difference equation
\begin{equation}
\label{slater1}
\left(p^2+\frac{1}{2}s-\epsilon\sqrt{s}\right)f_n(p)-\frac{1}{4}s
\left[f_n(p+2)+f_n(p-2)\right]=0\; ,
\end{equation}
where $p=kd/\pi$, $s=8U_0/E_R$ and $\epsilon=E_R\sqrt{s}/U_0$. 
In the low-lying energy bands $f_n(p)$  is a
slowly varying function of its argument, so that Eq. (\ref{slater1}) 
is reduced to the differential equation
\begin{equation}
\label{slater2}
-s{d^2\; f_n(p)\over dp^2}+(p^2-\epsilon\sqrt{s})f_n(p)=0\; .
\end{equation}
Evidently, this is the Schr\"odinger equation 
for a harmonic oscillator, with
energy levels $\epsilon_n=2n+1$
and eigenfunctions $f_n(\kappa)\propto \exp (-\kappa^2/2)H_n(\kappa)$   
where $\kappa=s^{-1/4}p$ and $H_n(\kappa)$ are the
Hermite polynomials.

The Wannier functions appropriate to the low-lying energy bands of
the 1D Mathieu
problem are, therefore, single-well orbitals with the same functional form
as for the harmonic
oscillator. From his study of the exact Wannier functions of the same
problem Slater [14] then
showed that these single-well orbitals provide an accurate decription of
the first five bands
under the restriction $s>16$\  i.e. $U_0>2E_R$. 
The condition of validity for the first
Wannier
function in Eq. (\ref{gauss}) 
can be estimated from Slater's numerical results as
$U_0>E_R/4$.

To extend this argument to the Wannier functions of an
interacting BEC, we view the mean-field interactions as
equivalent to a renormalization in the height of the external potential, by
an amount that we may
estimate as the average value of $g\phi^2(z)$ 
over a lattice cell. From the
normalization condition
satisfied by $\phi(z)$, 
in the case of repulsive interactions the height of the
periodic potential is then
{\it lowered} by an amount $g/d$. This leads to Eq. (\ref{slaterineq}) 
in the main text.

\section{Details on the on-site problem}

\subsection{On-site Bogolubov operators}
The operators $L_\pm$ entering the Bogolubov equations 
(\ref{bogoeqZa}) and (\ref{bogoeqZb}) 
can be split into an on-site part and a residual
part as  $L_\pm (z)=L^{(s)}_\pm (z)+\Delta L_\pm (z)$, where
\begin{equation}
\label{lsmeno}
L^{(s)}_-=-\nabla^2/2m+U^{(s)}_{p}+H_I^{(s)}-\mu_s\; ,
\end{equation} 
\begin{equation}
\label{lspiu}
L^{(s)}_+=L^{(s)}_-+2 H_I^{(s)}
\end{equation} 
and
\begin{equation}
\label{deltal}
\Delta L_- =\Delta U_p+\Delta H_I -\Delta\mu\; ,
\end{equation}
\begin{equation}
\label{deltalpiu}
\Delta L_+ =\Delta L_- +2\Delta H_I \; .
\end{equation}
In Eqs. (\ref{lsmeno}-\ref{deltalpiu}) $U^{(s)}_{p}$ is the
single-well
potential
$U^{(s)}_{p}=
U_p(z)\theta\left(1-4z^2/ d^2\right)$, 
$\theta(x)$ being the step function, 
$H_I^{(s)}$ is the on-site mean-field potential 
$H_I^{(s)}=gw_{0+}^2(z)$ and $\mu_s$ is the on-site chemical 
potential, while 
the  $\Delta$'s are
differences between
lattice and single-well values.
Namely,
\begin{equation}
\label{Deltaup}
\Delta U_p(z)=U_0\sin^2(kz)\theta\left({4z^2\over d^2}-1\right)
\end{equation}
and
\begin{equation}
\label{Deltahi}
\Delta H_I(z)=
g\left(|Z_{00+}(z)|^2-|w_{0+}(z)|^2\right)\; .
\end{equation}

\subsection{Variational calculation of Gaussian width}

A crucial quantity in the calculations reported in section 3.2
is the width $\sigma$ of
$\varphi_{0+}(z)$, since the integrals
entering Eq. (\ref{bandlow}) 
are determined by the overlap in adjacent wells. A
first estimate of $\sigma$ can be obtained 
from the curvature of the bottom of the
well: this determines
a frequency $\omega=(U_0E_R/2\hbar^2)^{1/2}$ such that  
$\sigma=(\hbar/m\omega)^{1/2}$. A more refined
variational estimate can be based on a method first proposed by Baym and 
Pethick [16], including
the effect of interactions in the condensate.

We construct the energy functional
\begin{equation}
\label{functional}
E[\phi]={\hbar^2\over 2m}\int dz\; \left|{d\phi(z)\over dz}\right|^2+
\int dz\; W(z)|\phi(z)|^2+\frac{1}{2}g\int dz\; |\phi(z)|^4
\end{equation}
and require that it be minimized by $\varphi_{0+}(z)$ when we set  
$W(z)=m\omega^2 z^2/2$. This yields the
energy function
\begin{equation}
\label{emin}
E(\sigma)={\hbar^2\over 4m\sigma^2}+{1\over
4}m\omega^2\sigma^2+{g\over 2\sigma\sqrt{2\pi}}\; ,
\end{equation}
which is minimized when $\sigma$  obeys the condition
\begin{equation}
\label{sigma}
\sigma^4=({\hbar\over m\omega})^2+{g\sigma\over
\sqrt{2\pi}m\omega^2}\; .
\end{equation}
Evidently, repulsive interactions broaden the Wannier function.
A similar estimate of $\sigma$ for the $n=1$ band states shows that
the role of the interactions is decreasing with increasing energy.

\end{document}